\documentclass[fleqn,usenatbib]{mnras}

\usepackage{newtxtext,newtxmath}

\usepackage[T1]{fontenc}

\DeclareRobustCommand{\VAN}[3]{#2}
\let\VANthebibliography\thebibliography
\def\thebibliography{\DeclareRobustCommand{\VAN}[3]{##3}\VANthebibliography}

\usepackage{graphicx}	\usepackage{amsmath}	\usepackage{lipsum}  
\newcommand{\solarradii}{$R_{\bigodot}$}

\title[Coherent deflection pattern in the solar wind]{Coherent deflection pattern and associated temperature enhancements in the near-Sun solar wind}

\author[R. Laker et al.]{
Ronan Laker,$^{1}$\thanks{E-mail: ronan.laker15@imperial.ac.uk}
T. S. Horbury,$^{1}$
L. D. Woodham,$^{1}$
S. D. Bale,$^{2,3}$
and L. Matteini$^{1}$
\\
$^{1}$Imperial College London, Blackett Laboratory, South Kensington, SW7 2AZ\\
$^{2}$Physics Department, University of California, Berkeley, CA 94720-7300, USA\\
$^{3}$Space Sciences Laboratory, University of California, Berkeley, CA 94720-7450, USA
}

\date{Accepted XXX. Received YYY; in original form ZZZ}

\pubyear{2023}

\begin{document}
\label{firstpage}
\pagerange{\pageref{firstpage}--\pageref{lastpage}}
\maketitle

\begin{abstract}

Measurements of transverse magnetic field and velocity components from Parker Solar Probe have revealed a coherent quasi-periodic pattern in the near-Sun solar wind.
As well as being Alfv\'enic and arc-polarised, these deflections were characterised by a consistent orientation and an increased proton core temperature, which was greater parallel to the magnetic field.
We show that switchbacks represent the largest deflections within this underlying structure, which is itself consistent with the expected outflow from interchange reconnection simulations.
Additionally, the spatial scale of the deflections was estimated to be around $1$\,Mm on the Sun, comparable to the jetting activity observed at coronal bright points within the base of coronal plumes.
Therefore, our results could represent the in situ signature of interchange reconnection from coronal bright points within plumes, complementing recent numerical and observational studies.
We also found a consistent relationship between the proton core temperature and magnetic field angle across the Parker Solar Probe encounters and discussed how such a persistent signature could be more indicative of an in situ mechanism creating a local increase in temperature.
In future, observations of minor ions, radio bursts and remote sensing images could help further establish the connection between reconnection events on the Sun and signatures in the solar wind.

\end{abstract}


\section{Introduction}

Switchbacks are Alfv\'enic deflections found throughout the solar wind \citep{Neugebauer2012,Horbury2018}, which have become increasingly important following the recent Parker Solar Probe (PSP) measurements \citep{Bale2019, Kasper2019}.
Their dominance of the magnetic field profile \citep{Bale2019}, along with their associated increases in bulk proton speed \citep{Matteini2014}, make them a potentially important component of the solar wind.
In addition, many of their proposed generation methods are linked to broader theories for how the solar wind is heated and accelerated \citep[e.g.][]{Fisk2020, Raouafi2023b}.
Therefore, by characterising the properties and behaviour of switchbacks, we can better understand the underlying mechanism occurring closer to the Sun and ultimately help reveal how the solar wind is formed.

While there has been much progress in switchback research over the first four years of the PSP mission, there is not yet enough evidence to form a consensus on whether switchbacks are created in situ, or launched by transient events on the Sun \citep[see ][for a review]{Raouafi2023}.
The simplest in situ mechanism relies on the expansion of the solar wind to grow Alfv\'enic fluctuations so that $\delta \vec{B}/|\vec{B}_0| > 1$ \citep{Barnes1974}, creating a reversal in the magnetic field as the fluctuation maintains a constant amplitude \citep{Mallet2021}. 
Expanding box simulations based on this concept \citep{Squire2020,Shoda2021} have demonstrated that switchbacks can naturally develop from initially small, random fluctuations.
However, such models struggle to generate the large number of switchbacks \citep{Bale2019,Shoda2021} and the preference for transverse deflection directions observed by PSP \citep{Laker2022,Fargette2022}.

In contrast, remote sensing observations have established that transient events (e.g. jets) are ubiquitous across the solar disc, regardless of the solar cycle \citep{McIntosh2014,Tian2014}.
Assuming that the products of these events can survive into the solar wind \citep{Landi2005,Tenerani2020}, they could supply enough energy to heat the solar wind \citep{DePontieu2007,Raouafi2023b} and create the switchbacks \citep{Neugebauer2012,Neugebauer2021}.
There are several ways in which the fundamental process of  magnetic reconnection on the Sun could create switchbacks in the solar wind, through the launching of: kinks in the magnetic field lines \citep{Fisk2020, Sterling2020}; flux ropes \citep{Drake2021} and fast mode waves \citep{Zank2020} or the jet material itself \citep{Neugebauer2021}.
While it is generally understood that modulation in switchback activity and alpha fraction (known as switchback patches) are the result of spatial modulation at scales consistent with supergranules/plumes near the Sun \citep{Fargette2021,Bale2021}, there has not been conclusive evidence linking switchbacks directly to a solar source.

Recently, there has been growing evidence in favour of interchange reconnection as the source of switchbacks.
Although coronal plumes are often seen as stable, ray-like structures \citep{DeForest2001}, detailed remote sensing observations have identified quasi-periodic jetting activity at their base \citep{Raouafi2008, Uritsky2021, Kumar2022, Raouafi2023b}.
It is thought that interchange reconnection at coronal bright points could be the source of this periodic signature, which was found to exist on the scale of minutes \citep{Uritsky2021, Kumar2022}.
Numerical simulations have also demonstrated that a single source of interchange reconnection can continuously generate Alfv\'enic fluctuations \citep{Wyper2022} with periodicity that matches the remote sensing observations \citep{Gannouni2023}.
In addition, \citet{Bale2023} found evidence of highly energetic protons and alphas ($10^{5}$\,eV), whose distribution was consistent with the output of an interchange reconnection simulation on the Sun.

Therefore, it appears that a compelling case for pervasive interchange reconnection is emerging from both remote sensing observations and numerical simulations.
Hence, it is anticipated that the solar wind displays a quasi-periodic signature, with potential heating, that would represent the `imprint of interchange reconnection' \citep{Wyper2022}.

The aim of this paper is to identify such a signature in the solar wind using in situ data from PSP at 13\solarradii{} from the Sun. 
In Section \ref{sec:deflections}, we first demonstrate that individual switchbacks are often part of a larger structure of coherent deflections in the solar wind.
We show that such a structure is characterised by quasi-periodic deflections in the transverse magnetic field components, which display a consistent orientation similar to the numerical simulations from \citet{Wyper2022}.
After estimating the corresponding spatial scale at the Sun, we further argue that the quasi-periodic deflections in the solar wind are consistent with jetting from coronal bright points observed at the base of plumes \citep{Uritsky2021,Kumar2022,Raouafi2023b}.
Section \ref{sec:temp} further investigates the proton core temperature inside each deflection, showing that there is a greater enhancement in the parallel, rather than perpendicular, direction.
While this is again suggestive of interchange reconnection, we further investigate the relationship between temperature and magnetic field angle, and discuss how this could be more consistent with an in situ heating mechanism.

\section{Coherent deflection signature}\label{sec:deflections}

At their simplest description, switchbacks are folds in the magnetic field, which appear as large rotations in the spacecraft data \citep{Bale2019}.
By applying a threshold to the magnetic field deflection angle (e.g. $>45^{\circ}$), previous studies have been able to uncover switchback characteristics by averaging over large numbers of switchbacks \citep{Mozer2020, DudokdeWit2020,Laker2021}.
While this is a valid way to investigate switchbacks, it is important to note how such a definition has influenced the way that switchbacks are conceptualised.
Under such a definition, switchbacks are considered to be a single deflection away from, and back to, the ambient magnetic field direction.
Therefore, with a constraint on the magnetic deflection angle, the internal structure of switchbacks is intrinsically tied to the definition of the switchbacks themselves.
This also means that a series of magnetic field deflections would be classified as several distinct switchbacks, rather than the sub-structure of some larger phenomenon.
Hence, in contrast, we directly examine the magnetic field deflections, rather than try to identify individual switchbacks.

\begin{figure}
    \centering
    \includegraphics[width=\columnwidth]{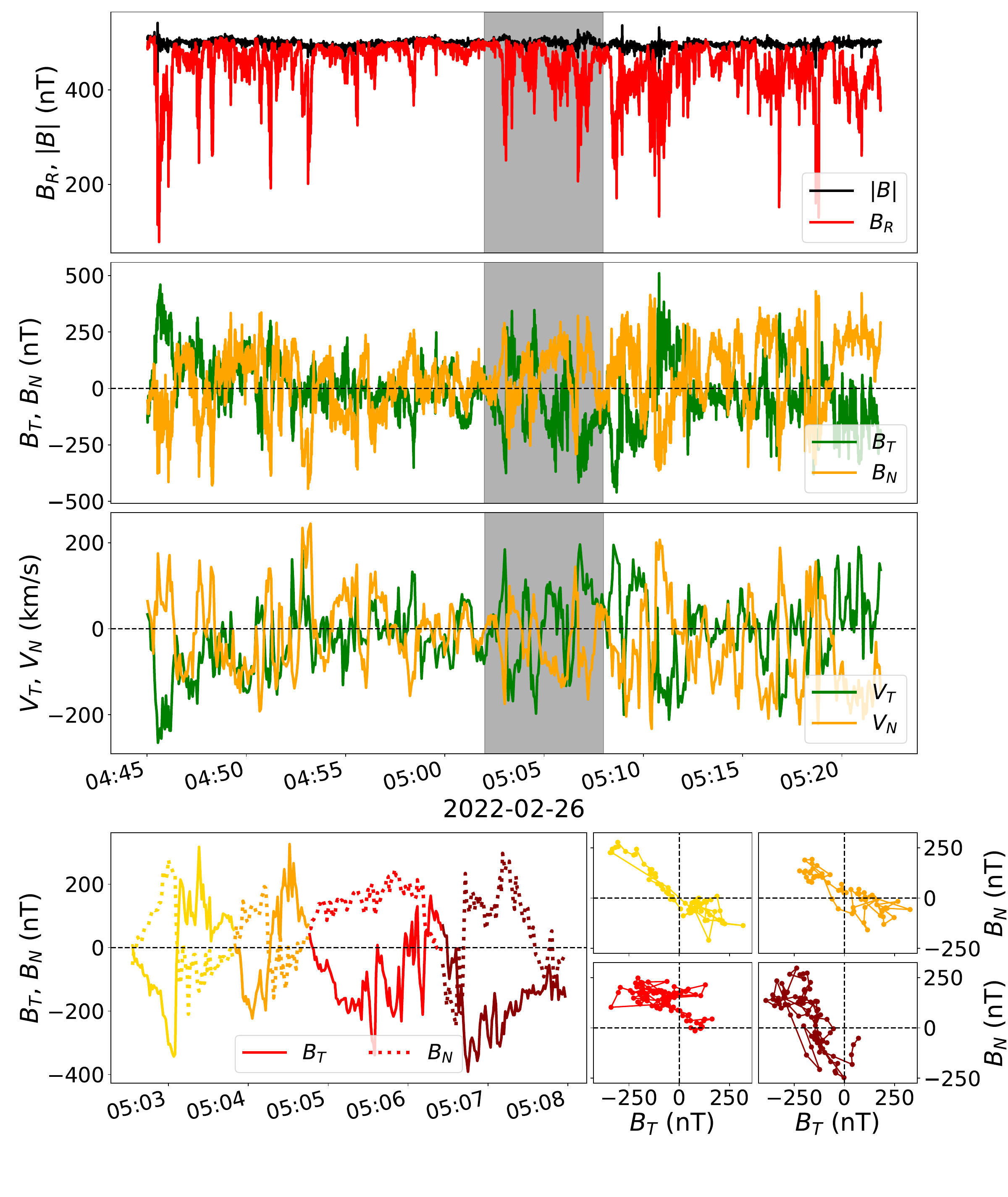}
    \caption[Example of anti-correlated deflections in $B_T$ and $B_N$]{Example patch of anti-correlated deflections in $B_T$ (green) and $B_N$ (orange). These deflections are Alfv\'enic fluctuations, as shown by a similar pattern in the velocity fluctuations (third panel). The hodograms for the highlighted region are shown in the lower panels, where several successive deflections, split by colour, all travel along the same arc.}
    \label{fig:anti_deflections}
\end{figure}

We use the 4 sample/cycle magnetic field measurements from FIELDS \citep[approximately 4.58 samples/second, ][]{Bale2016} up to PSP's eleventh encounter, which we refer to as E11.
Such data, an example period of which is shown in Fig. \ref{fig:anti_deflections}, represents the solar wind only $13$ solar radii (\solarradii{}) from the Sun.
When attempting to identify switchbacks by a threshold in deflection angle, dips in $B_R$\footnote{Radial-Tangential-Normal (RTN) coordinate system where $\vec{R}$ points from the Sun to the spacecraft, $\vec{N}$ is the component of the solar north direction perpendicular to $\vec{R}$, and $\vec{T}$ completes the right-handed set} are often used as a proxy for the presence of switchbacks \citep[e.g.][]{Bale2019, Kumar2023}, which appeared unremarkable for the period in Fig. \ref{fig:anti_deflections}.
However, by focusing on the transverse magnetic field components, $B_T$ and $B_N$ in the second panel, a more coherent pattern in the solar wind emerges.
Both the $B_T$ and $B_N$ components varied quasi-periodically, and appeared to be anti-correlated on the scale of a few minutes.
Much like previous switchback reports, these deflections were still Alfv\'enic and arc-polarised \citep{Horbury2020}, as demonstrated by the hodograms in the lower right panels.
However, successive deflections (depicted by the colour in the hodograms) are now seen to deflect back and forth with the same orientation, which manifests as the anti-correlation seen in the transverse components.
Such a quasi-periodic pattern can be found throughout the near-Sun solar wind, leading us to conclude that switchbacks, as they have been studied thus far, are part of this more fundamental structure in the solar wind. 
By applying a threshold to the deflection angle, only the most extreme parts of this pattern were sampled in previous studies, giving a glimpse of the underlying structure.

\begin{figure}
    \centering
    \includegraphics[width=\columnwidth]{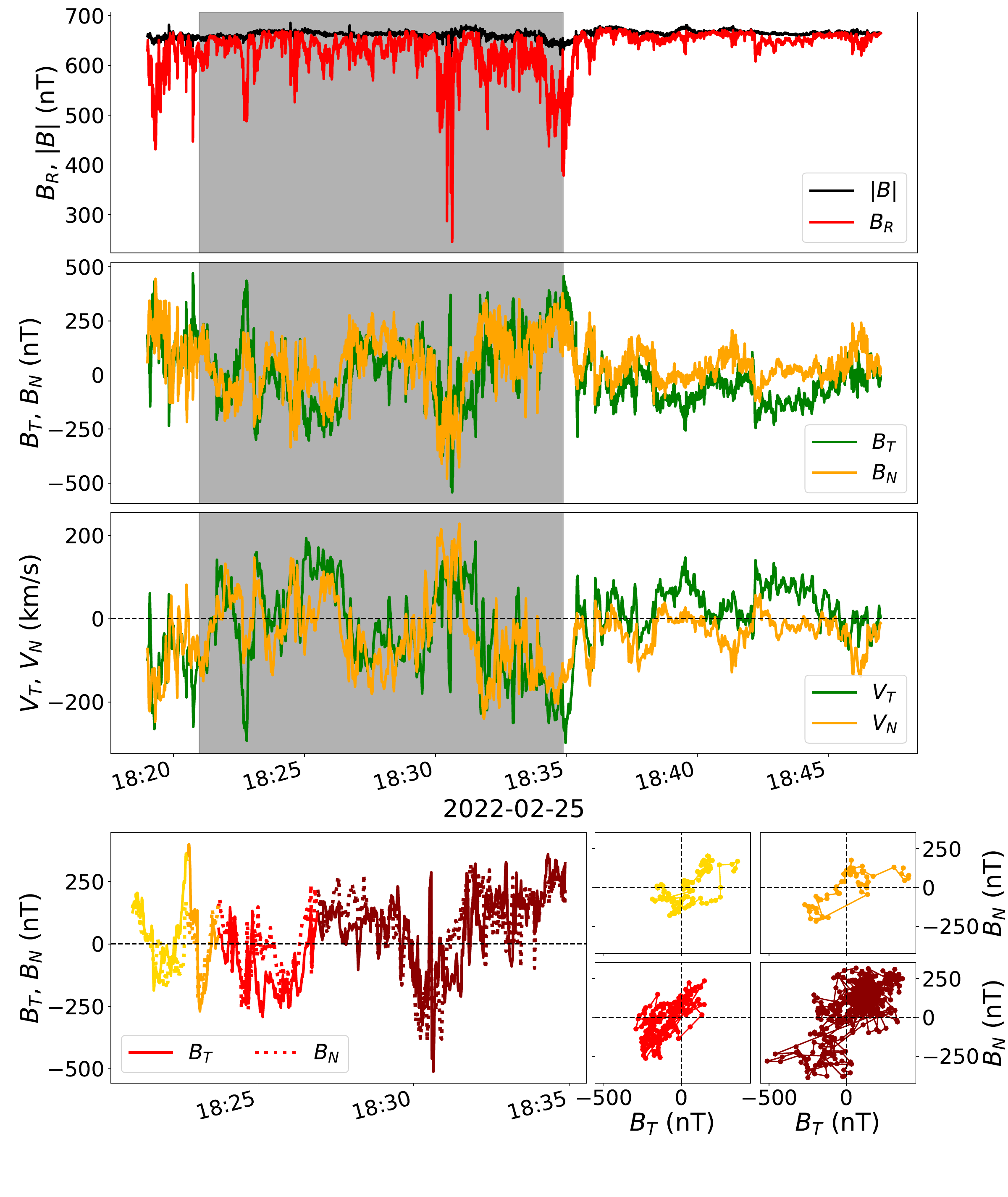}
    \caption[Example of correlated deflections in $B_T$ and $B_N$]{Example patch of correlated deflections in $B_T$ (green) and $B_N$ (orange). The hodograms for the highlighted grey region are shown below, where several successive deflections, split by colour, all travel along the same arc.}
    \label{fig:corr_deflections}
\end{figure}

While we found this anti-correlation of tangential components to be more common in the PSP data, Fig. \ref{fig:corr_deflections} demonstrates that there were also times when the transverse components were correlated.
The observed anti-/correlations were due to the deflection directions being oriented roughly -45\,$^{\circ}$ from the $B_N$ axis in Fig. \ref{fig:anti_deflections} and 45\,$^{\circ}$ for Fig. \ref{fig:corr_deflections} (where a positive angle is clockwise).
As a result, it appears that the magnetic field tends to deflect in either one of two orthogonal directions.
This matches the preferential deflection directions observed by analysing the switchback deflections over an entire PSP encounter in previous studies \citep{Fargette2022,Laker2022}.
In this sense, the quasi-periodic signature in Fig. \ref{fig:anti_deflections} and Fig. \ref{fig:corr_deflections} is the local manifestation (with a scale of minutes) of the statistical trend found over the whole encounter.

Our description of a quasi-periodic pattern with a consistent orientation is comparable to the expected outflow from the interchange reconnection occurring at an embedded bipole structure in \citet{Wyper2022}.
The authors found that a single source of interchange reconnection could create a continuous source of torsional Alfv\'en waves with the same sense of rotation.
These represent a pattern of alternating deflections in the solar wind, which are then sampled by the spacecraft to generate a quasi-periodic signal in the observed magnetic field.

\citet{Gannouni2023} found similar Alfv\'enic fluctuations with another interchange reconnection simulation, further showing that emergence of magnetic flux was enough to trigger the reconnection, rather than surface motions used by \citet{Wyper2022}.
The authors demonstrated that their model could periodically create velocity spikes every 19 minutes, which they concluded was consistent with the quasi-periodic jetting activity recently observed at the base of plumes \citep{Uritsky2021,Kumar2022}.
In this scenario, the quasi-periodic deflections at the spacecraft could be the direct consequence of repeated jetting observed on the Sun, with a period of $\sim 3 \to 5$\,minutes on the Sun \citep{Uritsky2021}, being reflected in the solar wind.
Unfortunately, due to spacecraft and plasma motions, it is not clear how a timescale at the spacecraft relates to the timescale seen on the Sun, meaning we can only investigate the spatial scale of the pattern.

\begin{figure}
    \centering
    \includegraphics[width=\columnwidth]{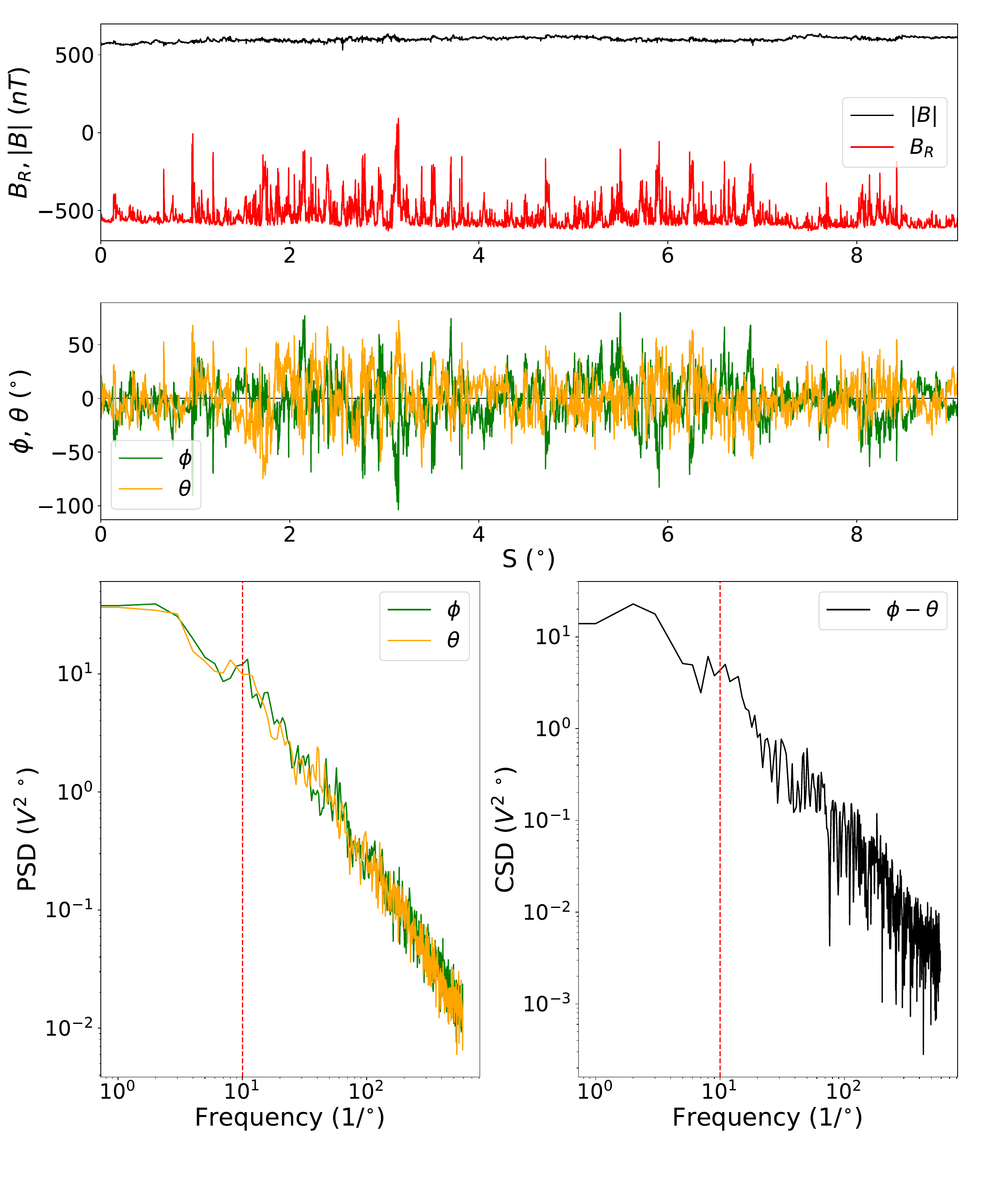}
    \caption{Example power and cross spectral density with respect to the angular displacement of the spacecraft for a 4-hour period in E10. There is enhanced power in both $\phi_B$ and $\theta_B$ around $0.1^{\circ}$ in the lower left panel. This scale also had an increased cross spectral density (lower right panel), indicating that they were co-moving.}
    \label{fig:example_E10_welch}
\end{figure}

Consequently, Fig. \ref{fig:example_E10_welch} displays the magnetic field angles ($\phi_B, \, \theta_B$) with respect to the angular displacement of the spacecraft's trajectory \citep[similar to ][]{Fargette2021} for a 3-hour period during E10.
The associated power spectra, calculated with Welch's method with a Hann window of 1$^{\circ}$, had enhanced power at 0.1$^{\circ}$ for both magnetic field angles and the cross-spectra.
This corresponds to a length scale of $\sim$\,16\,Mm at 13$R_{\bigodot}$ from the Sun, which equates to a length scale of $\sim$\,0.6\,Mm on the Sun \citep[using a typical super-radial expansion factor of $4$, ][]{Fargette2021, Uritsky2021}.
Such a scale is comparable to both granules \citep[$\sim$ 1\,Mm, ][]{Nordlund2009} and coronal bright points within plumes \citep[2"$\to$ 3" or 1.4 $\to$ 2.1\,Mm,][]{Kumar2022}, meaning it is plausible that the pattern of transverse deflections in the solar wind are the result of interchange reconnection at the base of plumes \citep{Raouafi2023b}. 

Therefore, it appears that the solar wind exhibits quasi-periodic variations within the larger scale modulation of switchback patches \citep{Fargette2021,Bale2021}, reflecting the substructure of plumelets/granules within plumes/supergranules at the Sun. 
This sub-patch structure was also found in previous studies that investigated periodicity in the magnetic deflection angle \citep{Fargette2021} and $V_R$ component \citep{Kumar2023}.
Crucially, we have now demonstrated that it is not just a periodic occurrence of switchbacks, but a coherent pattern of transverse magnetic field deflections that matches the expectations of numerical simulations \citep{Wyper2022,Gannouni2023}.

It is important to note that the existence of such a pattern of deflections is not exclusive to the interchange reconnection idea. 
Our observations of arc-polarised deflections and rotational discontinuities are similar to the previously reported `phase steepened Alfv\'en waves'  seen with Ulysses \citep{Tsurutani1999}. 
Numerical simulations have shown how Alfv\'en waves can develop into the arc-polarised structures and rotational discontinuities \citep{Barnes1974,Vasquez1996}, thereby not requiring a solar source.
However, it is less clear how such a process can account for the observed characteristic scale (order of minutes) and consistent orientation of the deflections \citep{Laker2022}.
While interchange reconnection can plausibly explain the preferential deflection directions as the result of open field lines being dragged over closed field loops \citep{Fisk2020, Fargette2022,Laker2022}, in situ mechanisms struggle to reconcile these observed switchback characteristics.
Interestingly, both \citet{Squire2022} and \citet{Johnston2022} demonstrated how an initially random distribution of deflections could be influenced by a non-radial magnetic field direction, such as the Parker spiral.
Alternatively, \citet{Vasquez1996} described how the modulation of the magnetic field could encourage the steepening of Alfv\'en waves in a particular direction.
In these scenarios, the orientation of the deflections would therefore reflect the configuration of the magnetic field closer to the Sun, e.g. the magnetic funnels proposed by \citet{Bale2021}.
Indeed, these in situ effects can also apply to fluctuations created by a solar source, therefore, as \citet{Wyper2022} suggested, the true answer is most likely a combination of mechanisms.

\section{Temperature enhancements in deflections}\label{sec:temp}

In addition to representing a more pristine state of the solar wind, the later PSP encounters ($\sim$\,13\solarradii{}) benefited from an enhanced transverse orbital speed, which increased the visibility of the proton core in the SPAN-Ai instrument \citep{Kasper2016}.

\begin{figure}
    \centering
    \includegraphics[width=\columnwidth]{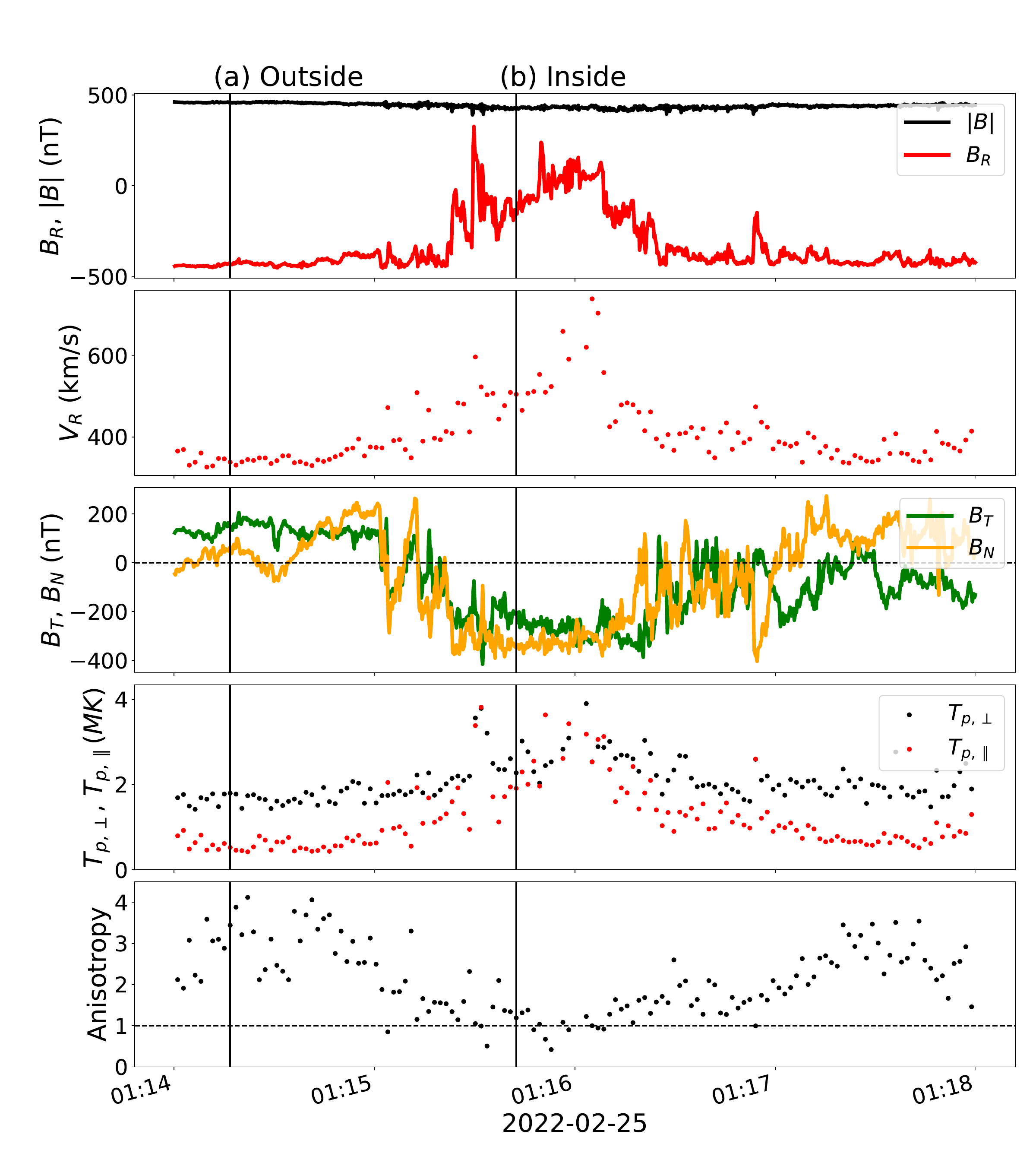}
    \caption{Example deflection from E11, demonstrating that both $T_{p,\parallel}$ and $T_{p,\perp}$ are enhanced inside, which reduces the anisotropy of the core to $\sim$1. Individual VDFs for inside (b) and outside (a) the deflection are shown in Fig. \ref{fig:example_VDF}.}
    \label{fig:example_T_increase}
\end{figure}

After applying the same method of \citet{Woodham2021} to the SPAN-Ai data, we plot the proton core temperature parallel ($T_{p, \parallel}$) and perpendicular ($T_{p, \perp}$) to the magnetic field during a deflection in Fig. \ref{fig:example_T_increase}.
Either side of the magnetic field deflection, the ambient solar wind displayed a typically high proton core temperature anisotropy ($T_{p, \perp}/T_{p, \parallel} \sim 3$) \citep{Marsch1982}.
However, as the magnetic field angle increased inside the deflection, both the $T_{p, \perp}$ and $T_{p, \parallel}$ increased, with the greater change in the parallel direction leading the proton core to become isotropic ($T_{p, \perp}/T_{p, \parallel} \sim 1$).

\begin{figure}
    \centering
    \includegraphics[width=\columnwidth]{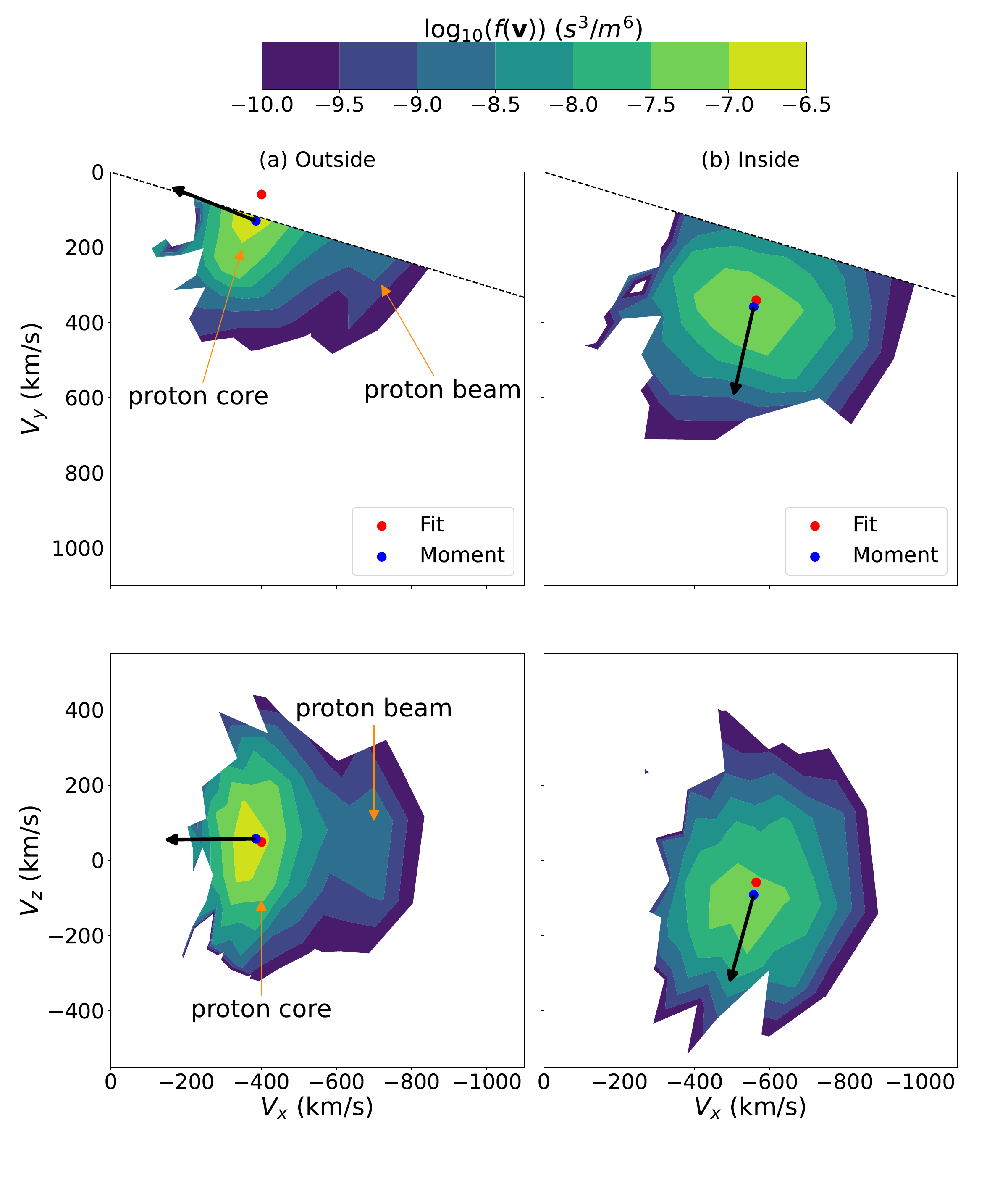}
    \caption{Example VDFs from inside (b) and outside (a) the deflection in Fig. \ref{fig:example_T_increase}. Plots are in the instrument frame, with the proton core and beam labelled as well as the velocity from the moment in blue, with the fitted velocity in red. The SPAN-Ai field of view is cut off at the dashed line, but as the magnetic field (black arrow) rotated in the deflection, the proton core became more visible. This is associated with an increased $T_{p,\perp}$ and $T_{p,\parallel}$. }
    \label{fig:example_VDF}
\end{figure}

To demonstrate that these temperature enhancements are not caused by an instrumental effect, we plot the corresponding velocity distribution functions (VDFs) for inside, Fig. \ref{fig:example_VDF}(b), and outside the deflection, Fig. \ref{fig:example_VDF}(a).
Again, Fig. \ref{fig:example_VDF}(a) demonstrates that the ambient solar wind had a typical anisotropic proton core and a proton beam population along the magnetic field direction (black arrow) \citep{Marsch1982}.
Fig. \ref{fig:example_VDF}(b) then shows how the proton core deflected with the magnetic field \citep{Matteini2015}, which was associated with an increase in  both $T_{p, \perp}$ and $T_{p, \parallel}$ leading to a reduction in the temperature anisotropy.
In this example, the plasma deflected towards $-V_T$, which rotated the proton core into the field of view (FoV) of SPAN-Ai, while rotating the proton beam in the opposite sense \citep{Matteini2014,Matteini2015}.
Conversely, a deflection towards $+V_T$ would reduce the visibility of the proton core, and lead to erroneous fits to the proton beam.
To avoid such times, we have enforced the condition that the angle of the solar wind from radial in the spacecraft frame is: 
\begin{equation}\label{eq:sw_angle}
    \tan^{-1} \left (\frac{V_{SW,\, T} - V_{SC, \,T}}{V_{SW, \,R}} \right ) < -5^{\circ},
\end{equation}
using velocity measurements derived from moments to the SPAN-Ai distributions.

\begin{figure}
    \centering
    \includegraphics[width=\columnwidth]{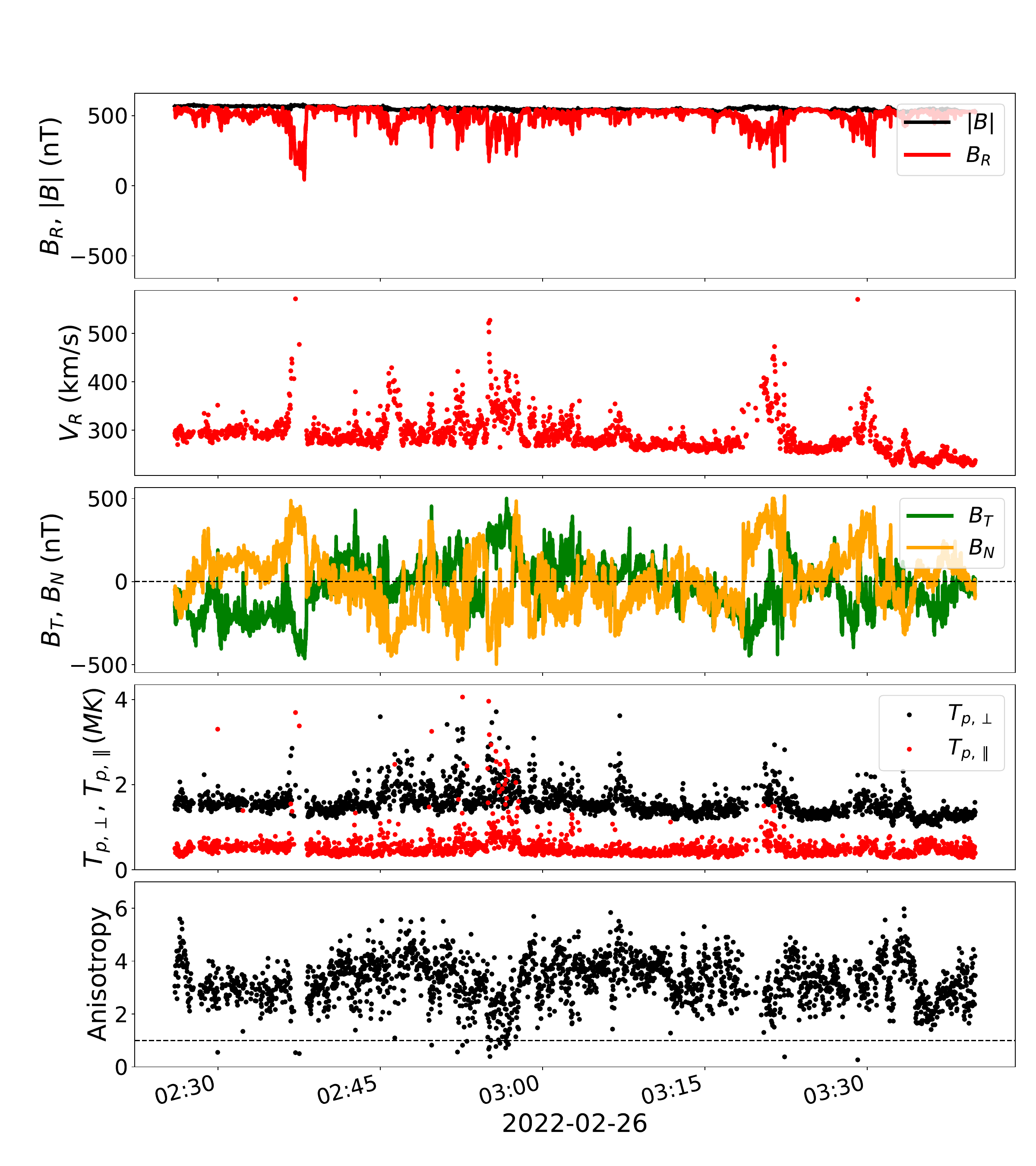}
    \caption{Example period of deflections in E11 which demonstrates that the increases in perpendicular and parallel temperature are also quasi-periodic.}
    \label{fig:E11_consistent_signature}
\end{figure}

Applying this condition to an extended period of solar wind, in Fig. \ref{fig:E11_consistent_signature}, demonstrates that the temperature enhancements were also quasi-periodic.
Similar enhancements in $T_{p, \parallel}$ were reported during E2 by \citet{Woodham2021}, with the authors concluding that `switchbacks [were] embedded within a larger-scale structure identified by distinct plasma signatures'.
In this view, switchbacks were first identified by their deflection angle, with larger scale groups of switchbacks (termed `patches') then being linked to an enhanced $T_{p, \parallel}$.
However, following the observations of a quasi-periodic pattern from Section \ref{sec:deflections}, we argue that these deflections, each with an enhanced temperature, should be identified first.
Any smaller-scale changes in the magnetic field would now be considered sub-structure of this quasi-periodic pattern, rather than individual structures themselves.
For example, despite there being several changes in the magnetic field direction during Fig. \ref{fig:example_T_increase}, there is only a single increase in temperature correlated with a magnetic deflection lasting several minutes.

Consequently, generation methods need to be able to explain the quasi-periodicity in both the magnetic field deflections and associated temperature increases.
Indeed, the greater increase of temperature in the parallel direction can naturally be explained by the injection of energy from magnetic reconnection, which has already been discussed as a potential source of the deflections in the previous section.
This connection is further supported by the recent observations of high energy protons ($\sim \, 10^{5}$\,eV) matching the expected power law from an interchange reconnection simulation \citep{Bale2023}. 
We note that the period of interest in \citet{Bale2023} also displays the quasi-periodic pattern from this study, therefore, it will be important to investigate if enhancements in high energy particles are also confined to these deflections.
Unfortunately, this may be technically challenging considering that the particle measurements were integrated over several hours for \citet{Bale2023}. 
It is important to note that while magnetic reconnection can create energised protons and Alfv\'en waves at comparable speeds, they will drift apart as the Alfv\'en speed increases in the solar corona. 
Therefore, it is unlikely that the energised protons would be in phase with the Alfv\'enic fluctuations by the time they have propagated to the spacecraft.
Instead, the spacecraft may be cutting through flux tubes that have been heated via repeated reconnection events, removing the need for the fluctuations to be in phase with the accelerated particles.

After removing the T-V relationship from different PSP encounters (see Appendix), Fig. \ref{fig:dT_theta} shows that there was a consistent relationship between the change in temperature ($\delta T_{p}$) and magnetic field angle from radial ($\theta_{RB}$).
This is more suggestive of an in situ heating mechanisms that is dependent on the angle of the magnetic field \citep{Woodham2021a, Opie2022}, rather than the presence of the magnetic deflection itself.

The coloured scatter points depict the average over 10$^{\circ}$ bins for each individual encounter.
Remarkably, the temperature profile was persistent across the different PSP encounters, with a much greater increase in $\delta T_{p, \parallel}$ than $\delta T_{p, \perp}$, which drove the proton core anisotropy below $1$ at $\sim$$ 75^{\circ}$, before reaching around $\sim$0.8 past $90^{\circ}$.
The temperature enhancements appeared to plateau above 75$^{\circ}$.
The average increase in temperature above this angle was more than twice as high for $\delta T_{p, \parallel}$, $1.46\pm0.56$\,MK, than $\delta T_{p, \perp}$, $0.55\pm0.28$\,MK as depicted by the black dashed line in Fig. \ref{fig:dT_theta}.

\begin{figure}
    \centering
    \includegraphics[width=\columnwidth]{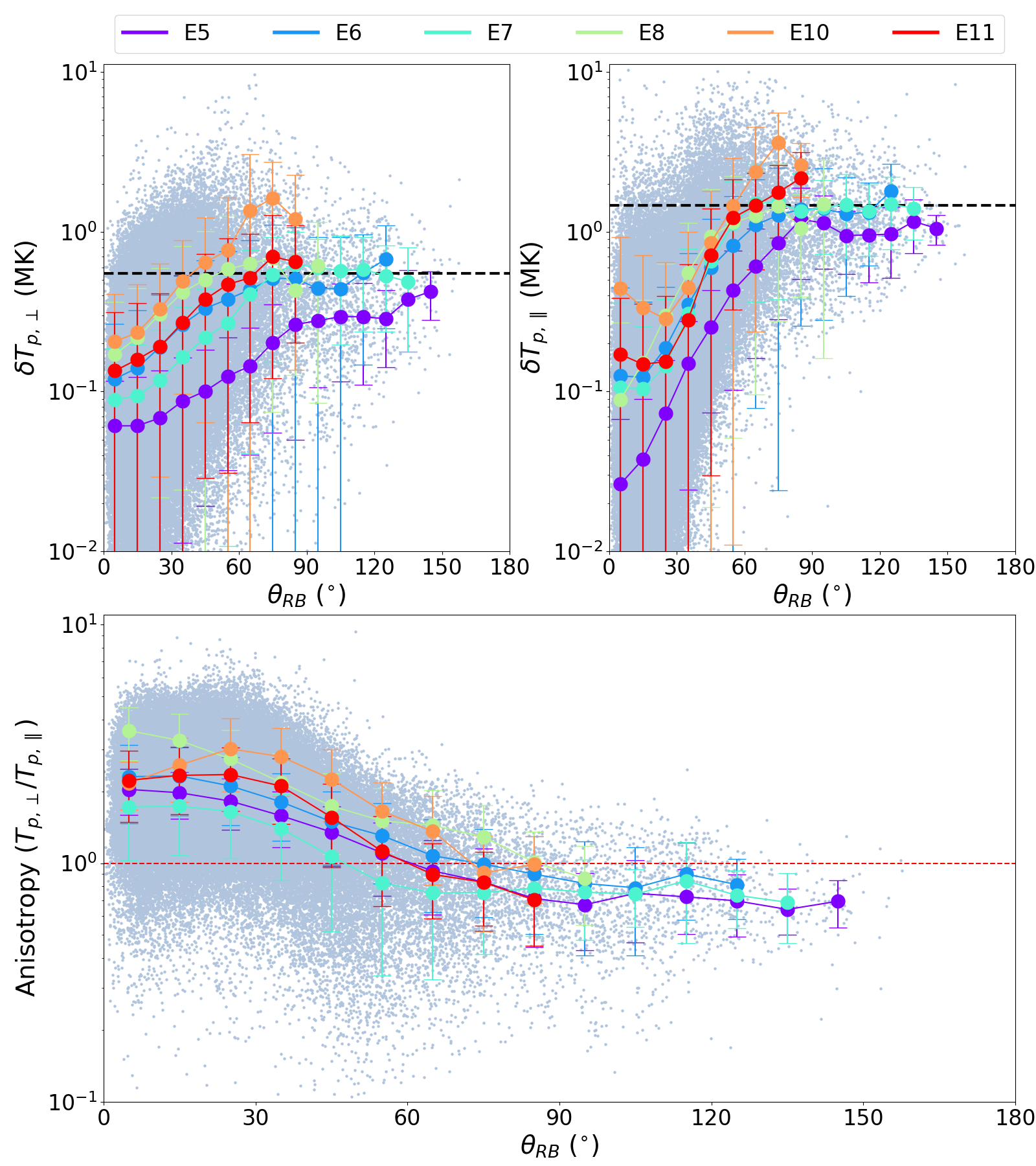}
    \caption[Temperature vs $\theta_{RB}$ for several encounters]{Relationship between $\delta T_{p,\perp}$, $\delta T_{p,\parallel}$ and anisotropy with $\theta_{RB}$ after the removal of the background T-V relationship. Mean average and standard deviation are shown for $10^{\circ}$ bins of $\theta_{RB}$ for each encounter. The dashed black line represents the mean temperature increase for angles $\ge 75^{\circ}$. This value was more than twice as high for $\delta T_{p, \parallel}$, $1.46\pm0.56$\,MK, than $\delta T_{p, \perp}$, $0.55\pm0.28$\,MK}
    \label{fig:dT_theta}
\end{figure}

Such a result appears to be in contention with an earlier study from \citet{Woolley2020} that reported $T_{p,\parallel}$ was equal at $\theta_{RB} = 0^{\circ}$ and $180^{\circ}$ with the Solar Probe Cup \cite[SPC,][]{Case2020}.
While it would be preferable to study the same switchbacks as \citet{Woolley2020}, the orbital speed of PSP was too low for SPAN-Ai to study the proton core during the first encounter.
Conversely, the lower value of $\delta \vec{B} / \vec{B}$ meant that the magnetic field rarely rotated near $180^{\circ}$ in the later encounters (e.g. E10 and E11), even though there was a higher orbital speed.
We were unable to identify a full reversal switchback with reliable data from both instruments to compare their performance.
Consequently, there remains a need for future investigations into the plasma properties at full reversals.

The consistent rate of temperature increase across the PSP encounters does suggest that the temperature enhancements are created by some in situ mechanism that is dependent on the magnetic field angle \citep{Opie2022}.
For example, using Wind spacecraft data at 1\,AU, \citet{Woodham2021a} interpreted changes in $T_{p, \parallel}$ and $T_{p, \perp}$ with $\theta_{RB}$ as turbulent dissipation through kinetic Alfv\'en waves and Alfv\'en ion cyclotron waves, respectively.
Similarly, \citet{DAmicis2019} also demonstrated a strong positive correlation with $T_{p, \perp}$ and $\theta_{RB}$ using Wind, which they linked to ion-cyclotron resonance.
Interestingly, \citet{DAmicis2019} did not find any correlation between $T_{p, \parallel}$ and $\theta_{RB}$ at 1\,AU even though it was more prevalent than perpendicular heating at PSP \citep{Woodham2021}.
In contrast, Helios measurements from 0.3\,AU showed no increase in temperature in either direction \citep{Horbury2018}, although this may be due to an instrumental effect \citep{DAmicis2019}

Therefore, it seems that the increases in parallel temperature are confined to the near-Sun environment, while ion-cyclotron heating exists out to 1\,AU \citep{Perrone2019,Hollweg2002}.
Perhaps this is due to the $T_{p,\parallel}$ enhancement being related to: the phase steepening of the Alfv\'en waves \citep{Vasquez1998,Gonzalez2021}; shear induced effects \citep{DelSarto2018}; or the parametric instability that requires compressive fluctuations \citep{Tu1995,Hollweg2002} that could only exist close to the Sun.

No matter the mechanism, it will be important to understand if the temperature enhancements within the deflections can affect the bulk temperature of the solar wind, or if they are just local modulations of the plasma conditions.
Given that the solar wind is well known to exhibit perpendicular heating with distance from the Sun \citep{Hellinger2011,Perrone2019}, and the deflections have primarily $T_{p, \parallel}$ increases, it does not seem likely that they heat the solar wind directly.
This is also reflected in the fact that there is a stronger T-V relationship for $T_{p, \perp}$ in both Table \ref{tab:TV_results}  and \citet{Perrone2019a}.
Therefore, the coherent deflections may represent the tracers of a heating process closer to the Sun, such as interchange reconnection, without heating the wind directly \citep{Raouafi2023b}.

\section{Conclusions}\label{sec:conclusions}

In this study, we have presented evidence for a coherent pattern in the transverse magnetic field, which was associated with increased proton core temperature in both parallel and perpendicular directions.
This quasi-periodic pattern was characterised by successive deflections having a consistent orientation, with a greater increase in parallel temperature that reduced the anisotropy of the proton core.

As well as matching the expected outflow from interchange reconnection simulations \citep{Wyper2022,Gannouni2023}, the spatial size of the deflections was estimated as around 1\,Mm on the Sun, making them comparable to scale of coronal bright points that exhibit jetting activity at the base of coronal plumes \citep{Raouafi2008, Uritsky2021,Kumar2022,Kumar2023}.
Therefore, we argue that the quasi-periodic pattern of deflections in the solar wind, within larger scale switchback patches, mirrors the plumelet sub-structure observed within coronal plumes on the Sun.
Such a connection can further explain the greater increase in parallel temperature, and has also been cited as the source of high energy particles in the solar wind \citep{Bale2023}.

After removing the larger scale T-V relationship, we demonstrated that the proton core temperature increases associated with the deflections were remarkably consistent with magnetic field angle across the PSP encounters.
The strong link between temperature and magnetic field angle does imply that an underlying mechanism is responsible for the characteristic temperature increases, such as: heating through the phase steepening of Alfv\'en waves \citep{Vasquez1998,Gonzalez2021}; turbulent dissipation \citep{Woodham2021a}; shear induced effects \citep{DelSarto2018}; or wave-particle interactions \citep{DAmicis2019,Opie2022}.
So while we believe our results do support the growing evidence from remote sensing and simulation domains, we cannot yet rule out in situ mechanisms from our results alone.

Although the enhancements in $T_{p, \parallel}$ are clear at PSP with our study and \citet{Woodham2021}, this relationship was not found at Helios \citep{Horbury2018} or Wind \citep{DAmicis2019}.
Consequently, it is not yet known how the deflections merge and contribute to the bulk speed and temperature of the solar wind.
This issue can be directly addressed by Solar Orbiter, which was in direct radial alignment with PSP during its 11$^{\textrm{th}}$ encounter in February 2022.
In future, a coordinated effort of remote and in situ data could track reconnection outflows out to spacecraft distances, providing conclusive evidence for the source of the solar wind.

\section*{Acknowledgements}

RL was supported by an Imperial College President's Scholarship and ST/W001071/1, TSH by ST/W001071/1 and LDW by STFC ST/S000364/1. The SWEAP and FIELDS teams acknowledge support from NASA contract NNN06AA01C. This work has made use of the open source and free community-developed space physics packages HelioPy \citep{Stansby2021a} and SpiceyPy \citep{Annex2021}.

\section*{Data Availability}

The data used in this study are available at the NASA Space Physics Data Facility (SPDF): https://spdf.gsfc.nasa.gov.

\bibliographystyle{mnras}
\bibliography{library.bib}

\appendix

\subsection{Removing the T-V relationship}

\begin{figure}
    \centering
    \includegraphics[width=\columnwidth]{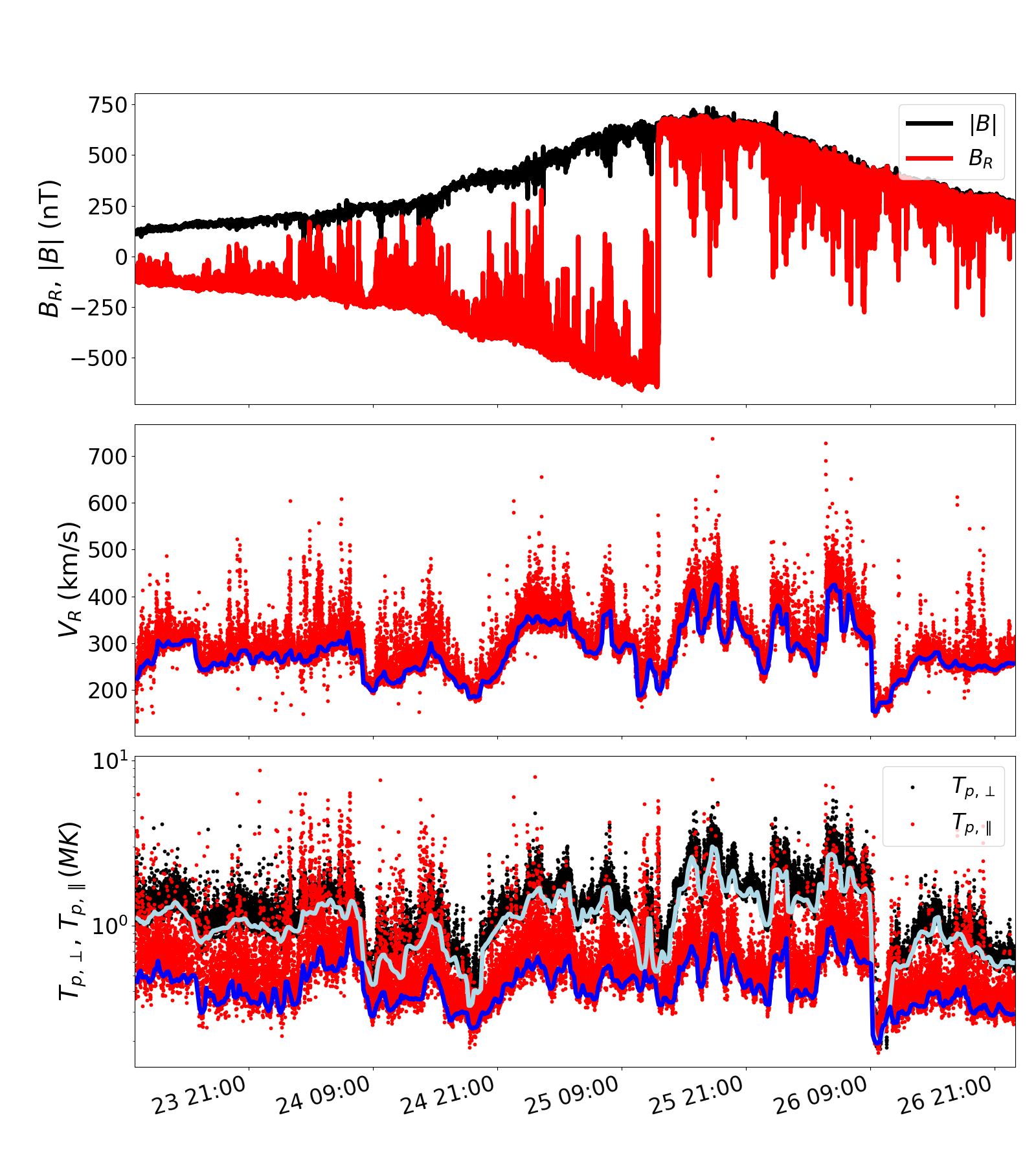}
    \caption{Plasma parameters for a period of E11 where the light blue ($T_{p,\perp}$) and dark blue ($T_{p,\parallel}$) lines show the 30-minute rolling averaging used to get the baseline variations in temperature and speed found in Table \ref{tab:TV_results}.}
    \label{fig:E11_rolled}
\end{figure}

To compare the temperature enhancements across many switchbacks, we first removed the well established T-V relationship from the different solar wind streams \citep{Neugebauer1966,Burlaga1970,Elliott2012}.
The baseline temperature and speed was estimated by taking the 10\% percentile for a 30\,minute rolling window (with step size of 10\,minutes), as shown in Fig. \ref{fig:E11_rolled}.
For each PSP encounter, we identified a period of Alfv\'enic solar wind and fitted a linear relationship between the baseline temperatures, $ \left< T_{p, \perp} \right>$ and $\left< T_{p, \parallel} \right>$, and speed, $\left< V \right>$, with the results found in Table \ref{tab:TV_results}.
While the exact relationship changed between encounters, we found that $ \left< T_{p, \perp} \right>$ had a significantly stronger relationship with $\left< V \right>$ than $ \left< T_{p, \parallel} \right>$, matching the expectations from \citet{Perrone2019}.

\begin{table*}
    \centering
    \begin{tabular}{c|c|c|c|c|c}
    Encounter & Start & End &  Min. Distance  &  $ \left< T_{p, \perp} \right>$ gradient   & $ \left< T_{p, \parallel} \right>$ gradient \\ 
    & & & from Sun (\solarradii{}) & ($10^{3}$Ks/km) & ($10^{3}$Ks/km) \\ \hline
    5 & 2020-06-05 10:30 & 2020-06-07 02:54 & 28.1 & $3.71 \pm 0.17$ & $0.75 \pm 0.13$ \\
    6 & 2020-09-25 19:59 & 2020-09-29 16:34 & 20.4 & $3.52 \pm 0.20$ & $0.52 \pm 0.10$ \\
    7 & 2021-01-15 07:34 & 2021-01-18 09:34 & 20.4 & $3.73 \pm 0.18$ & $0.99 \pm 0.10$ \\
    8 & 2021-04-26 21:00 & 2021-04-28 13:00 & 19.0 & $2.78 \pm 0.42$ & $-0.15 \pm 0.12$ \\
    10 & 2021-11-19 16:00 & 2021-11-21 16:00 & 13.3 & $4.92 \pm 0.17$ & $1.12 \pm 0.08$ \\
    11 & 2022-02-23 10:00 & 2022-02-26 23:00 & 13.3 & $8.80 \pm 0.17$ & $2.15 \pm 0.07$ \\
    \end{tabular}\caption{Periods of the PSP data used in this study, along with the gradients of large scale T-V relationship. The measured T-V relationship seemed to match the predictions of \citet{Perrone2019}, this is beyond the scope of this paper and will be left for a future study.}\label{tab:TV_results}
\end{table*}

\bsp	\label{lastpage}
\end{document}